\documentclass[12pt,preprint]{aastex}
\usepackage{amssymb}
\usepackage{amsmath}
\usepackage{epstopdf}
\usepackage{natbib}
\usepackage[colorlinks, citecolor=blue]{hyperref}
\usepackage{hyperref}
\usepackage[all]{hypcap}

\begin{document}

\title{Superluminous Supernovae Powered by Magnetars: Late-time Light Curves and Hard Emission Leakage}
\author{S. Q. Wang\altaffilmark{1,2}, L. J. Wang\altaffilmark{1,2}, Z. G. Dai\altaffilmark{1,2} and X. F. Wu\altaffilmark{3,4,5}}
\affil{\altaffilmark{1}School of Astronomy and Space Science, Nanjing University, Nanjing 210093, China; dzg@nju.edu.cn}
\affil{\altaffilmark{2}Key Laboratory of Modern Astronomy and Astrophysics (Nanjing University), Ministry of Education, China}
\affil{\altaffilmark{3}Purple Mountain Observatory, Chinese Academy of Sciences, Nanjing, 210008, China}
\affil{\altaffilmark{4}Chinese Center for Antarctic Astronomy, Chinese Academy of Sciences, Nanjing, 210008, China}
\affil{\altaffilmark{5}Joint Center for Particle Nuclear Physics and Cosmology of Purple Mountain
Observatory-Nanjing University, Chinese Academy of Sciences, Nanjing 210008, China}

\begin{abstract}
Recently, researches performed by two groups have revealed that the magnetar spin-down energy injection model with full energy trapping
can explain the early-time light curves of SN~2010gx, SN~2013dg, LSQ12dlf, SSS120810 and CSS121015,
but fails to fit the late-time light curves of these Superluminous Supernovae (SLSNe).
These results imply that the original magnetar-powered model is challenged in explaining these SLSNe.
Our paper aims to simultaneously explain both the early- and late-time data/upper limits
 by considering the leakage of hard emissions.
We incorporate quantitatively the leakage effect into the original magnetar-powered model and derive a new semi-analytical equation.
Comparing the light curves reproduced by our revised magnetar-powered model to the observed data and/or upper limits of these five SLSNe,
 we found that the late-time light curves reproduced by our semi-analytical equation are in good agreement with
 the late-time observed data and/or upper limits of SN~2010gx, CSS121015, SN~2013dg and LSQ12dlf and
  the late-time excess of SSS120810, indicating that the magnetar-powered model might be responsible for these
 SLSNe and that the gamma ray and X-ray leakage are unavoidable when the hard photons were down-Comptonized to softer photons.
To determine the details of the leakage effect and unveil the nature of SLSNe,
 more high quality bolometric light curves and spectra of SLSNe are required.
\end{abstract}


\section{Introduction}

In recent years, several Wide Field Optical Transient projects, e.g., the Palomar Transient Factory \citep[PTF;][]{Law2009,Rau2009},
the Panoramic Survey Telescope \& Rapid Response System \citep[Pan-STARRS1, PS1;][]{Kai2010,Ton2012},
the Catalina Real-time Transient Survey \citep[CRTS;][]{Dra2009}, La Silla Quest Supernova Survey \citep[LSQ;][]{Bal2013} have discovered a class of supernovae (SNe)
with peak luminosities $30-100$ times more luminous than normal SNe.
Accordingly, they were termed ``Superluminous Supernovae (SLSNe)" \citep{Gal2012,Qui2012} and are
distinct from their low luminosity cousins.

Determining the heating sources for SLSNe and clarifying the explosion mechanisms of their progenitors would be of outstanding importance for the
 theory of supernovae and of stellar evolution.
While observational studies of both light curves and spectra of SLSNe have been largely pursued in the last decade
 \citep[see][and references therein]{Gal2012},
major progress has been made in understanding their nature. However, it is still not completely clear how SLSNe with
 a variety of characters in their light curves and spectra are powered.

The $^{56}$Ni cascade decay (radioactivity-powered) model \citep{Col1969,Col1980,Arn1982,Woos1989,Val2008}
is the most ``natural" model accounting for most SNe and has been strongly supported by a great amount of unambiguous observations.
However, for very luminous SNe, the masses of $^{56}$Ni needed are so large that we have to consider the
``pair instability SNe (PISNe)" \citep{Bar1967,Rak1967,Heg2002,Heg2003,Cha2012a,Wha2013,Des2013,Cha2013b,Koz2014,Che2014a,Che2014b,Che2014c}
 which are very rare in low-redshift universe \footnote{The mass loss rate of a massive star scales
 with $(Z/Z_{\odot})^{0.5-0.8}$ \citep{Kud1987,Lei1992,Vin2001}, where $Z$ and $Z_{\odot}$ are the metallicity
of the massive stars and solar metallicity respectively. Nearly all of massive stars formed in low-redshift universe have metallicities $Z\gtrsim 0.1Z_{\odot}$
and lose most of their masses before their death. Hence the main-sequence masses of these stars must be very large,
 $M_{\rm ZAMS}$ = 140 to 260 $M_{\odot}$ for $Z=0$ \citep{Heg2002} and $\gtrsim 500 M_{\odot}$ for $Z\gtrsim 0.1Z_{\odot}$ \citep{Che2014c}.
 However, very massive stars occupy the top of the initial mass function (IMF) and are very rare in low-redshift universe.
The unique low-redshift PISN candidate to date is SN~2007bi ($z = 0.123$) \citep{Gal2009}, which has also been
 explained by other scenarios \citep[e.g.,][]{Mor2010,Kas2010,Des2012}.}.
Other challenges posed on the $^{56}$Ni cascade decay model
  are that more massive $^{56}$Ni needs more massive ejecta resulting in broad theoretical light curves
 and that the $^{56}$Co decay tails have rather shallow decline rate, both of which are not always consistent with observations \citep{Mae2007,Nich2013}.

The ejecta-circumstellar medium (CSM) interaction model \citep{Che1982,Che1994,Chu1994,Che2011,Cha2012}
is also often used to explain luminous and superluminous Type IIn SNe whose optical spectra show narrow and/or intermediate-width H$\alpha$ emission
 lines indicative of interactions between the SN ejecta and CSM which was ejected from the progenitor just prior to the genuine SN explosion.
\citet{Cha2013a} have applied their semi-analytical model \citep{Cha2012} to the observed light curves of a dozen
SLSNe and suggested that both hydrogen-deficient (Type I) SLSNe and hydrogen-rich (Type II) SLSNe can be powered by ejecta-CSM
interactions.

The rotational energy of a rapidly rotating, strongly magnetized pulsar forming after the core collapse of a massive star
 could significantly affect the evolution and emission of an outflow surrounding the star. It has been proposed \citep[see][]{Dai1998a,Dai1998b,Zhang2001,Dai2004,Yu2007,Dal2011,Dai2012}
that the rotational energy of a nascent magnetar can be injected into the relativistic outflow
of a gamma-ray burst (GRB) by magnetic dipole radiation and could lead to the shallow decay of a GRB afterglow,
 which is very consistent with observations by the $Swift$ satellite \citep{Zhang2007}.

The magnetar-powered scenario has also been introduced to explain the light curves of some SNe or SLSNe.
\citet{Mae2007} suggested that a magnetar with surface magnetic field $B \sim 10^{14-15}$ G and initial spin
period $P_0 \sim $10~ms can explain the second peak and late-time decline rate of the light curve of SN~2005bf;
\citet{Kas2010} and \citet{Woos2010} calculated the theoretical light curves powered by millisecond magnetars;
\citet{Inse2013} employed the semi-analytical magnetar-powered model to fit the
light curves of six SLSNe Ic (PTF10hgi, SN~2011ke, PTF11rks, SN~2011kf, SN~2012il and SN~2010gx) with a variety of
feasible physical values for magnetars, while the $^{56}$Ni-powered model needs unrealistic amount of $^{56}$Ni to explain these SLSNe and therefore can be excluded.
Hence, they concluded that the magnetar-powered model can provides a viable explanation for all of the SLSNe Ic.
The successful fit for another SLSN Ic, PTF12dam \citep{Nich2013}, with magnetar-powered model supported the conclusion raised by
\citet{Inse2013}. This so-called ``magnetar-powered model" has attracted more and more attentions
because it can explain many bright and/or sophisticated light curves of a variety of SNe while the traditional radioactivity-powered model cannot.

Recently, \citet{Nich2014} fitted the light curves of three SLSNe Ic (LSQ12dlf, SSS120810, SN~2013dg) and one SLSN IIn (CSS121015)
using the $^{56}$Ni cascade decay model, magnetar-powered model, and ejecta-CSM interaction model, respectively.
Their results excluded the $^{56}$Ni cascade decay model and found that the late-time light curves reproduced by the magnetar-powered model
are brighter than the observations, posing a challenge for the magnetar-powered model.
Hence, they suggested either that the magnetar-powered model fails to explain the light curves of those SLSNe, or that there exists X-ray leakage in
ejecta of these SLSNe \footnote{Note, however, that a similar difficulty was already encountered
 when \citet{Inse2013} fitted the light curve of SN~2010gx, although it has not been pointed out by the authors.}. They have alternatively fitted the data of these SLSNe using the ejecta-CSM interaction model and obtained rather good agreements.
Comparing these results, they proposed that these two models are valid in explaining SLSNe, without particularly favoring either. This conclusion is not only different from \citet{Inse2013} and \citet{Nich2013} who favor the
magnetar-powered model but also from \citet{Cha2013a} and \citet{Ben2014} who favor the interaction model.
Therefore, the energy generation mechanisms responsible for SLSNe Ic are still under debate.

Our major goals in this paper are to check the possibility that the magnetar-powered model can be responsible for some SLSNe Ic or IIn and to
 confirm quantitatively the leakage effect in the expanding ejecta.
For these purposes, we try to incorporate the leakage effect into the original magnetar-powered model, supposing that there exists
 gamma ray and X-ray leakage in SNe ejecta after energy was injected to the ejecta.

This paper is organized as follows: in Sect. \ref{sec:model} we revise the magnetar-powered model by introducing the leakage factor into the original model.
We fit the light curves of SN~2010gx, CSS121015, SN~2013dg, LSQ12dlf and SSS120810 using our revised magnetar-powered model in Sect. \ref{sec:results}.
Finally, discussions and conclusions are presented in Sect. \ref{sec:conclusion}.

\section{A Semi-Analytical Model and Its Revisions}
\label{sec:model}

We first derive the semi-analytical equation of SN light curves powered by magnetars in
this section and revise it by introducing the leakage effect.

Using Equations (30)-(33), (36), (47-48) derived by \citet{Arn1982} and neglecting the
 $(E_{th,0}/t_{0})e^{(t/\tau_{m})^2+(R_{0}/v\tau_{m})(t/\tau_{m})}$ term
 (where $t_{0}$ is the diffusion timescale, $E_{th,0}$ is the initial total thermal energy and proportional to $R_{0}^3$ and $T_{00}^4$,
  $R_{0}\ll 10^{14}$~cm is the initial radius of the progenitor, $T_{00}$
is the initial temperature of the center of the ejecta),
we get the photospheric luminosity of SNe powered
 by a variety of energy sources (radioactive $^{56}$Ni decay, magnetar spinning-down, etc),
\begin{eqnarray}
L(t)~=&&\frac{2}{\tau_{m}}e^{-\left(\frac{t^{2}}{\tau_{m}^{2}}+\frac{2R_{0}t}{v\tau_{m}^{2}}\right)}~
\int_0^t e^{\left(\frac{t'^{2}}{\tau_{m}^{2}}+\frac{2R_{0}t'}{v\tau_{m}^{2}}\right)}  \nonumber\\
     &&\times \left(\frac{R_{0}}{v\tau_{m}}+\frac{t'}{\tau_{m}}\right)L_{\rm inp}(t')dt' ,
\label{equ:lum}
\end{eqnarray}
where $\tau_{m}$ is the effective light curve timescale, $L_{\rm inp}(t)$ is the power function of the generalised power source.
 Using Equation (10) in \citet{Cha2012}, $\tau_{m}$ can be written as
\begin{eqnarray}
\tau_{m}~&=&\left(\frac{10\kappa M_{\rm ej}}{3\beta vc}\right)^{1/2} \nonumber\\
         &\simeq& \left(\frac{10\kappa M_{\rm ej}}{3\times 13.8\times vc}\right)^{1/2} \nonumber\\
         &=& 0.5\left(\frac{\kappa M_{\rm ej}}{vc}\right)^{1/2}
\label{equ:tau_m}
\end{eqnarray}
where $\kappa$, $M_{\rm ej}$, $v$ and $c$ are the Thomson electron scattering opacity,
 the ejecta mass, the expansion velocity of the ejecta and the speed of light in a vacuum, respectively.
$\beta \simeq 13.8$ is a constant that accounts for the density distribution of the ejecta.

Although the rotational energy dissipation mechanism of the magnetars is ambiguous,
we can expect that most of the rotational energy released is likely to be Poynting-flux-dominated wind \citep{Met2014}.
When the high energy leptons and/or baryons in the magnetar wind hit the SN ejecta,
X-ray and gamma ray photons can be generated and propagate in the ejecta.
Therefore it is reasonable to assume that most of the spin-down energy can be converted to X-ray and gamma ray photons.

In the magnetar-powered model, adopting the \textit{ansatz} that all the spin-down energy released
can be converted to the heating energy of surrounding SN ejecta, the input power for the ejecta is equal to the magnetic dipole radiation power,
\begin{equation}
L_{\rm inp({\rm mag})}(t) = \frac{E_{p}}{\tau_{p}} \frac{1}{(1+t/\tau_{p})^{2}},
\label{equ:input-mag}
\end{equation}
where $E_{p}\simeq~({1}/{2})I_{\rm NS}\Omega_{\rm NS}^2$ is the rotational energy of a magnetar,
 $I_{\rm NS}~=(2/5)M_{\rm NS}R_{\rm NS}^{2}$ is the moment of inertia of a magnetar
whose canonical value can be set to $10^{45}$~g~cm$^{2}$ \citep{Kas2010}. Thus,
$E_{p}\simeq 2\times 10^{52}~{\rm erg}\left({M_{\rm NS}}/{1.4~M_\odot}\right)\left({P_{\rm NS}}/{1~{\rm ms}}\right)^{-2}\left({R_{\rm NS}}/{10~{\rm km}}\right)^2$,
$\tau_{p}~=~{6I_{\rm NS} c^3}/{B^2R_{\rm NS}^6\Omega^2} = 1.3~B_{14}^{-2}P_{10}^2~ {\rm yr}$ is the spin-down timescale of the magnetar \citep{Kas2010},
 in which the magnetar dissipates most of its rotational energy into the expanding ejecta. Here $M_{\rm NS}$, $P_{\rm NS}$ and $R_{\rm NS}$ are the mass, rotational period and radius of the magnetar, respectively. The parameters have been scaled to typical values $B_{14}~=~B/10^{14}~{\rm G}$ and $P_{10}~=~P/10~{\rm ms}$ \citep{Kas2010}.

Most of previous researches that fit observed data with the magnetar-powered model neglected hard (gamma ray and X-ray)
 photon leakage at all epochs \footnote{After our manuscript was completed, we noted that \citet{Che2014} also independently
 took the leakage effect into account in their fitting to PTF12dam.},
whereas in reality a portion of hard photons can escape from the ejecta while other photons can be down-Comptonized to UV-optical-IR photons and diffuse to the photosphere, especially after the transition from the photospheric phase to the nebular phase.

Observations for SN~1987A \citep{Sun1987,Mat1988} and SN~2014J \citep{Chu2014,Dieh2014} have revealed that a large amount
of gamma rays and X-rays escape from their ejecta at later times.
Recently, \citet{Kot2013} have explored the effect of
 mildly magnetized millisecond pulsars ($B = 10^{13}$~G) on SN remnants and demonstrated that
 the hard emissions could generate bright ($10^{43}-10^{44}$ erg~s$^{-1}$) TeV gamma ray emissions
 and a milder X-ray peak ($10^{40}-10^{42}$ erg~s$^{-1}$) could appear several hundred days after the core-collapse supernova (CCSN) explosion.

Discussions of the details of the gamma ray and X-ray emission leakage are beyond the scope of this paper.
We then introduce the leakage factor $ke^{-\tau_{\gamma}}$ and the trapping factor $(1-ke^{-\tau_{\gamma}})$ in the magnetar-powered model
 to represent gamma/X-ray leakage and trapping rate, where $k \simeq 1$ is a dimensionless parameter \footnote{In previous researches, \citet{Woos1989}
adopted $k~=~1$ to fit the late-time light curve of SN~1987A, while \citet{Sol2000} and \citet{Sol2002}
adopted $k~=~0.965$ to fit the late-time light curve of SN~1998bw.}.
$\tau_{\gamma}$ is the optical depth of the ejecta to gamma rays and can be written as $\tau_{\gamma}=At^{-2}$ \citep{Clo1997,Cha2009,Cha2012}.
The larger $\tau_{\gamma}$ is, the larger the trapping rate is and the smaller the leakage rate is.

The density profile of the ejecta can be represented by a power law, $\rho(r, t)\propto [r(t)]^{-\eta}$.
Provided that the SN ejecta has a uniform density distribution ($\eta=0$, $\rho$=$\rho(t)$, $M_{\rm ej}=(4/3) \pi \rho R^3$) and the
expansion velocity is constant, we have $E_{\rm k}=(3/10)M_{\rm ej}v^2$ and $A$
 can be written as \footnote{When $\eta>0$, the general expression of $A$ ($=T_0^2$) can be found from Equations (4) and (5) of \citet{Clo1997}, but
 the clerical error in the right side of their Equation (4) must be corrected by removing the ``$-$".}
\begin{eqnarray}
A&=&\frac{9\kappa_{\gamma} M_{\rm ej}^2}{40\pi E_{\rm k}} \nonumber \\
 &=&\frac{3\kappa_{\gamma} M_{\rm ej}}{4\pi v^2 }\nonumber \\
 &=&4.748 \times 10^{13}\frac{\kappa_{\gamma,0.1}M_{\rm ej,\odot}}{v_{9}^2} {\rm s}^2,
\label{equ:leakage}
\end{eqnarray}
where $v_{9}=v/10^9$~cm~s$^{-1}$, $M_{\rm ej,\odot}=M_{\rm ej}/M_\odot$ and $\kappa_{\gamma,0.1}=\kappa_{\gamma}/0.1$ cm$^2$~g$^{-1}$,
$\kappa_{\gamma}$ is the effective gamma ray opacity.
 The typical value of $A$ for normal core collapse SNe (CCSNe) is between $10^{13}$~s$^2$ and $10^{15}$~s$^2$.

Using Equations (\ref{equ:lum}), (\ref{equ:input-mag}), (\ref{equ:leakage}) and let $k=1$, $R_{0}~\rightarrow~0$,
the photospheric luminosity function in the magnetar-powered model with leakage can be written as
\begin{eqnarray}
L_{\rm mag}(t)=&&\frac{2E_{p}}{\tau_{p}\tau_{m}} e^{-\left(\frac{t}{\tau_{m}}\right)^{2}}
\left(1-e^{-At^{-2}}\right)\nonumber\\
&&\times \int_{0}^{t}\frac{1}{(1+t'/\tau_{p})^{2}} e^{\left(\frac{t'}{\tau_{m}}\right)^{2}} \frac{t'}{\tau_{m}}dt'.
\label{equ:L-mag}
\end{eqnarray}

We refer to the model corresponding to Equation (\ref{equ:L-mag}) as the ``revised magnetar-powered model".
When $A~\rightarrow~\infty$, the leakage can be neglected and the model can be called as the ``original magnetar-powered model".

Therefore, the free parameters of our model are $M_{\rm ej}$, $v$ and $\kappa$ (which determine $\tau_{m}$; Equation (\ref{equ:tau_m})),
 $B$ and $P$ (which determine $E_{p}$ and $\tau_{p}$), and $A$.
The velocity $v$, if measured, can be fixed at its measured value and cannot be regarded as a free parameter.

If we assume $\eta=0$, $A$ can be calculated from $M_{\rm ej}$, $v$ and $\kappa_{\gamma}$ using Equation (\ref{equ:leakage}).
Furthermore, as has been shown by \citep{Kot2013}, for gamma rays $\kappa_{\gamma} \simeq \kappa$.
However, as the density profile and $\kappa_{\gamma}$ are quite uncertain \footnote{$\eta$ can vary between approximately 0 and 14 \citep{Che1982}, $\kappa_{\gamma}$ can vary between approximately 0.01 and 0.2 cm$^2$~g$^{-1}$ while the effective X-ray opacity $\kappa_{X}$
  can vary between approximately 0.2 and $10^4$ cm$^2$~g$^{-1}$ \citep[see][Figure 8]{Kot2013}.},
 the value of $A$ may be quite different from that calculated by Equation (\ref{equ:leakage}).
We therefore also investigate the effect of varying $A$ to represent cases where $\eta>0$ and/or $\kappa_{\gamma} \neq \kappa$.

\section{Results}
\label{sec:results}

In this section, we try to fit the observed data of SN~2010gx, CSS121015, SN~2013dg, LSQ12dlf and SSS120810 using our revised magnetar-powered model.
Our fits follow the researches performed by \citet{Inse2013} and \citet{Nich2014} whose fit parameters for these SLSNe are listed in Table \ref{parameters}
and are all denoted as ``Model A".
The common feature of the theoretical light curves reproduced by ``Model A" is that the early-time theoretical light curves were well fitted (except for LSQ12dlf) while the
late-time theoretical light curves are brighter than the observed luminosities.
i.e., the original magnetar-powered model fails to fit their late-time light curves.
An excess in the theoretical light curve reproduced for SN~2013dg is not obvious, but may still exist (see a discussion in section \ref{subsec:SN2013dg}).

We now employ our revised magnetar-powered model to fit the observed data and upper limits of aforementioned SLSNe
 and explore whether the late-time excess can be eliminated or decreased.

Based on the same parameters as \citet{Inse2013} and \citet{Nich2014} (except for SN~2013dg and LSQ12dlf),
we incorporated the leakage effect into the model, i.e., the values of $A$ are not set to be infinity,
 according to Equation (\ref{equ:leakage}), we calculate the values of $A$ for these five SLSNe, see Table \ref{parameters}.
The light curves reproduced both by the original and revised models were shown in Figures \ref{fig1} - \ref{fig5}.

\subsection{SN~2010gx}
\label{subsec:SN2010gx}

\citet{Cha2013a} have fitted the early-time ($\lesssim 100$~days) data of SN~2010gx using the
 $^{56}$Ni-, magnetar- and interaction-powered models and have excluded the $^{56}$Ni-powered model as $^{56}$Ni mass needed (12.97 $M_\odot$)
 is larger than the ejecta mass (5.17 $M_\odot$), see their Table 3.
\citet{Inse2013} fitted the early- and late-time data using the original magnetar-powered model. Their fit have late-time excess in
 the theoretical light curve.

Therefore, we try to fit the light curve using our revised magnetar-powered model.
If $\eta~=~0$ and $\kappa_{\gamma} = \kappa$, the value of $A$ for SN~2010gx is $2.212 \times 10^{14}$~s$^2$.
Our fit for SN~2010gx shows that our revised model can eliminates the late-time excess in the theoretical light curve reproduced by the
 original magnetar-powered model and well fits the whole light curve of SN~2010gx, see Figure \ref{fig1},
 indicating that our magnetar-powered model can be responsible to SN~2010gx and
 that the leakage is unavoidable in the expanding ejecta, although we cannot
 verify the precise amount of leakage due to the lack of later time data.

\subsection{CSS121015}
\label{subsec:CSS121015}

CSS121015 is a unique Type IIn SLSN in our sample here and can be well explained by the CSM interaction model \citep{Nich2014}.
 If the leakage was neglected ($\kappa_{\gamma}=\infty$),
 the late-time light curve reproduced by the magnetar-powered model is brighter than observations.

Here we fit the light curve using our revised magnetar-powered model.
If $\eta~=~0$ and $\kappa_{\gamma}=\kappa$, the value of $A$ for CSS121015 is $1.612 \times 10^{14}$~s$^2$.
The light curve reproduced by another set of parameters (see Model B for CSS121015 in Table \ref{parameters})
shows that the late-time excess in the magnetar-powered model can also be canceled by introducing the leakage effect,
see Figure \ref{fig2}.

\subsection{SN~2013dg}
\label{subsec:SN2013dg}

For SN~2013dg, since the size of the host galaxy contribution to the photometry at $\sim 200$~d is unknown \citep{Nich2014},
 the SN may be fainter than the data.

If the flux of SN~2013dg far exceeds the flux of the host at $\sim 200$~days, the later can be neglected and the fit is good enough.
However, if the flux of the host is comparable to the flux of SN~2013dg, the values of the last two data should be reduced by approximately 50\%.
So the late-time excess may still exist in the original magnetar-powered model.

\citet{Nich2014} did not fit the two uncertain data points with their CSM model.
Although the luminosities of the last two data must be fainter than the real luminosities, it is unreasonable to exclude them from analysis.
 So the CSM-interaction model cannot solely account for the late-time light curve of SN~2013dg.
In order to fit the last two data, substantial $^{56}$Ni is required and the parameters must be significantly changed.

Considering the difficulties mentioned above, we attempt to employ our revised magnetar-powered model to fit the light curve of SN~2013dg.
We first calculate the ``canonical" value $A=1.583 \times 10^{14}$~s$^2$ (corresponding the case of $\eta~=~0$ and $\kappa_{\gamma}=\kappa$) using Model B
 and find that the late-time theoretical luminosities are dimmer than the data without subtracting the contribution from the host galaxy.
If we regard the luminosities at $\sim 200$~days as precise ones, i.e., the contribution from the host is neglected,
 we find that when $A~=~6.5 \times 10^{14}$~s$^2$ (then $\kappa_{\gamma} \neq \kappa$)
  the reproduced light curve can get a better match for the last two data than the original magnetar-powered model.

As $A$ is a free parameter when $\eta~>~0$ and/or $\kappa_{\gamma} \neq \kappa$,
 the late-time light curve can vary continually from one situation to another, so we always can choose
a reasonable value of $A$ to fit the real tail of SN~2013dg as long as the values with subtracting the host contributions are in a reasonable
range.

The light curves reproduced by other sets of parameters (see Models B and C for SN~2013dg in Table \ref{parameters})
 are shown in Figure \ref{fig3}.

The advantage of our revised magnetar-powered model for explaining the SN~2013dg is that
 it can simultaneously explain both the early- and late-time data/upper limits
without introducing any amount of $^{56}$Ni which is needed for the CSM-interaction model,
 while the original magnetar-powered model can only fits the SN-dominated data
  and the CSM-interaction model has to discard the last two data.

\subsection{LSQ12dlf}
\label{subsec:LSQ12dlf}

Two problems exist in the fit for LSQ12dlf performed by \citet{Nich2014} using the magnetar-powered model: the earlier luminosities reproduced far
 exceed the observed upper limit and the late-time luminosity is also brighter than the last observed datum, see
 Figure 9 in \citet{Nich2014} and ``Model A" light curve in our Figure \ref{fig4}.

We adjust both the parameters of the magnetar and the explosion time and hence avoid the confliction in the early epoch and reduce the
late-phase excess. The new light curve (``Model B" light curve) is narrower than the light curve reproduced by the original magnetar-powered model.

In order to eliminate the late-time excess in the theoretical light curve,
 we take into account the leakage effect. If $\eta~=~0$ and $\kappa_{\gamma} = \kappa$, due to the large ejecta mass and low expansion velocity, the value of $A$ for LSQ12dlf is rather large, $7.62 \times 10^{14}$~s$^2$, resulting in a minor discrepancy between the ``Model C" light curve and ``Model B" light curve, so the leakage cannot effectively reduce the late time excess.
If $\eta~>~0$ and/or $\kappa_{\gamma} \neq \kappa$, we take the value of $A$ to be $3.50 \times 10^{14}$~s$^2$ (``Model D"), then the late-time excess can be completely eliminated.

\subsection{SSS120810}
\label{subsec:SSS120810}

Two main difficulties in the fits for SSS120810 are the rebrightening (``bump") in the late-time ($\sim$ 100 days after the explosion) light curve
and the late-time excess. We first deal only with the latter.

If $\kappa_{\gamma} = \kappa=0.20$~cm$^2$ g$^{-1}$, we calculate the value of $A$ to be $2.28 \times 10^{14}$~s$^2$ (``Model B") and obtain a light curve in which the excess can be partly canceled.

If $\eta~>~0$ and/or $\kappa_{\gamma} \neq \kappa$, we take the value of $A$ to be $1.50 \times 10^{14}$~s$^2$ (``Model C"),
 then the late-time excess is further eliminated but does not vanish completely and the early-time light curve
 begin to deviate slightly from the observations.

Although the later phase excess of light curve reproduced by the revised magnetar-powered model
 for SSS120810 cannot be completely eliminated by late-time leakage,
 the problem of excess was alleviated and the discrepancies between the data and the theoretical values are in a reasonable range, see Figure \ref{fig5}.

The rebrightening of SSS120810 cannot be explained by any model solely.
\citet{Nich2014} proposed that the energy resulting in the rebrightening might come from interactions between multiple CSM shells
 or breakout of a magnetar ionization front \citep{Met2014}.

We also suggest that the light curve of SSS120810 can be decomposed into two components. In
addition to the magnetar rotational energy injection which powers a monotonically declined
 light curve after the peak, something else, e.g., the thermal energy deposited by the shock wave in the oxygen shell \citep{Kle2014},
  may contribute to the formation of the bump and compensate the deviation in our ``Model C"
  light curve during $\sim 100-130$ days after the explosion. Detailed analysis and
  modeling for the bump are beyond the scope of this paper and we no longer discuss them here.

\section{Discussions and Conclusions}
\label{sec:conclusion}

In the past five years, many authors have demonstrated that many SLSNe Ic and II/IIn cannot be powered by radioactivity solely.
Alternatively, light curves reproduced by the magnetar spin-down model and the ejecta-CSM interaction model can both
well mimic the observed data of these SLSNe, indicating that most SLSNe might be powered by newly born magnetars or ejecta-CSM interactions.

\citet{Cha2013a} argued that almost all SLSNe can be powered by ejecta-CSM interactions.
Furthermore, \cite{Nich2014} have already demonstrated that CSS121015, LSQ12dlf, SSS120810 and SN~2013dg,
 which are difficult to be explained by the original magnetar-powered model due to
 their late-time excess in the theoretical light curves, can be explained by the ejecta-CSM interaction model,
 giving us a hint that some SLSNe Ic might
 be powered by Hydrogen- and Helium-poor CSM interactions.

However, no spectrum of a SLSN Ic so far has shown narrow emission lines indicative of the interaction of the ejecta with the CSM.
 This fact seems to suggest that the ejecta-CSM interaction model is difficult to be responsible for SLSNe Ic
  \footnote{But it should be noted that \citet{Ben-A2014} have demonstrated that SN~2010mb
 is a SN Ic which has a light curve that cannot be powered by $^{56}$Ni/$^{56}$Co but has a signature of interactions, indicating
 that it is a SN Ic powered by CSM-interaction and there might exist SLSNe Ic which can be explained by the CSM-interaction model as this
  model predicts a wide range of peak luminosities.}, although \citet{Cha2013a}
 argued that the absence of this character does not necessarily imply that the CSM interaction model is invalid for SLSNe Ic.

Besides the observational constraint aforementioned, the difficulties encountered by the interaction model for SLSNe Ic must also
be discussed from theoretical perspectives.
The first difficulty is that the formation mechanisms of
 massive ($M_{\rm CSM}\gtrsim 2 M_\odot$) H- and He-poor CSM are ambiguous.
 One of the main mechanisms explaining unusual high
 mass loss rate is the so-called ``pulsational pair-instability (PPI)" \citep{Heg2003,Woos2007,Pas2008,Chu2009,Cha2012b},
  which involves a very massive ($100 M_\odot \lesssim M_{\rm ZAMS} \lesssim140$$M_\odot$) star that can experience
 a series of eruptions caused by $e^{+}e^{-}$ pair generation
 and consequent $\nu\bar{\nu}$ pair creation but the star cannot be destroyed completely until it explodes as a genuine CCSN.
 The progenitors of SNe/SLSNe Ic are linked to the stars without H and He envelopes which are so compact that their outer materials are difficult to be ejected
 by PPI or other possible mechanisms. Combining the fact that no SLSNe presented emission lines, we argued that the CSM interaction model is difficult to account for at least some of SLSNe Ic.
  Another assumption adopted in the CSM interaction model employed by \citet{Cha2012}, \citet{Cha2013a} and \citet{Nich2014} is that the photospheres of SLSNe is stationary, whereas in reality the photospheres must expand due to the law of conservation of momentum.

The magnetar-powered model does not encounter the difficulties mentioned above.
 \citet{Kou1998} have demonstrated that magnetars with surface magnetic fields $\sim 10^{14-15}$~G
may constitute 10\% of nascent neutron stars \citep[see also][]{Kou1994,Par1995,Lyn1998},
 therefore, \citet{Woos2010} believed that the birth of rapidly rotating magnetars (``millisecond magnetars")
 is commonplace \footnote{Theoretical studies \citep[e.g.,][]{Met2007,Buc2008,Kom2008,Met2011} suggest
  that nascent millisecond proto-magnetars can explain long duration gamma ray bursts (LGRBs) and hyper-energetic
SNe (Hypernovae, HNe). Nevertheless, there is no clear observational evidence in favor of \textit{millisecond} proto-neutron stars with $P_{0} \lesssim 5$~ms
\citep[e.g.,][]{Vin2006}. Millisecond pulsars observed to date
  are all old, recycling (i.e. spun-up) pulsars rather than nascent ones.
   Further observations and theoretical analysis are needed to clarify this issue.}.
 Furthermore, the stationary photosphere assumption is not adopted in the magnetar-powered model.
Successful fits to the observed light curves of PTF10hgi, SN~2011kf, SN~2012il and PTF12dam using the magnetar-powered model \citep{Inse2013,Nich2013}
 have confirmed the validity of the original magnetar-powered model in explaining the luminosity evolution of the SLSNe Ic.

Yet the original magnetar-powered model is not a complete model since it neglects the hard emission leakage.
In reality, although the ejecta of SNe are opaque for gamma-ray and X-ray photon released from the magnetars at early times and these hard emission leakage are not significant, when the fast-expanding ejecta becomes more and more transparent for radiation at advanced epochs the hard photons would be pretty easy to escape from the ejecta and cannot be neglected.
So it is not surprising that the main problem encountered by the original magnetar-powered model is the late-time excess in the theoretical light curves of some SLSNe
(e.g., SN~2010gx \citep{Inse2013}, CSS121015, LSQ12dlf, SSS120810 and SN~2013dg \citep{Nich2014}).

This significant challenge confronted by the original magnetar-powered model motivates us to explore possible revisions for the original magnetar-powered model.
 We therefore consider the leakage effect and find that the late-time light curves derived from our revised magnetar-powered model are in good agreement with observed data and upper limits, indicating not only that the magnetars can be responsible to some SLSNe but also that the gamma-ray and X-ray leakage are inevitable when these photons were down-Comptonized to UV-optical-NIR photons and diffuse out of the ejecta.

Accordingly, we argue that it is not necessary to apply the ejecta-CSM interaction model in explaining the observed light curves for some SLSNe Ic (e.g., SN~2010gx, SSS120810, LSQ12dlf and SN~2013dg). Furthermore, we find that SLSN IIn CSS121015 which can be explained by ejecta-CSM interaction also can be powered by our revised magnetar-powered model, indicating
that H-rich SLSNe powered by magnetars may exist in the universe.
Hence we can expect that most SLSNe Ic can be powered by nascent millisecond magnetars and a (minor) portion of SLSNe II can be powered by
nascent millisecond magnetars or in rare cases by recycling millisecond magnetars \citep{Bar2011}.

\begin{figure}[htbp]
\begin{center}
   \includegraphics[width=1.0\textwidth,angle=0]{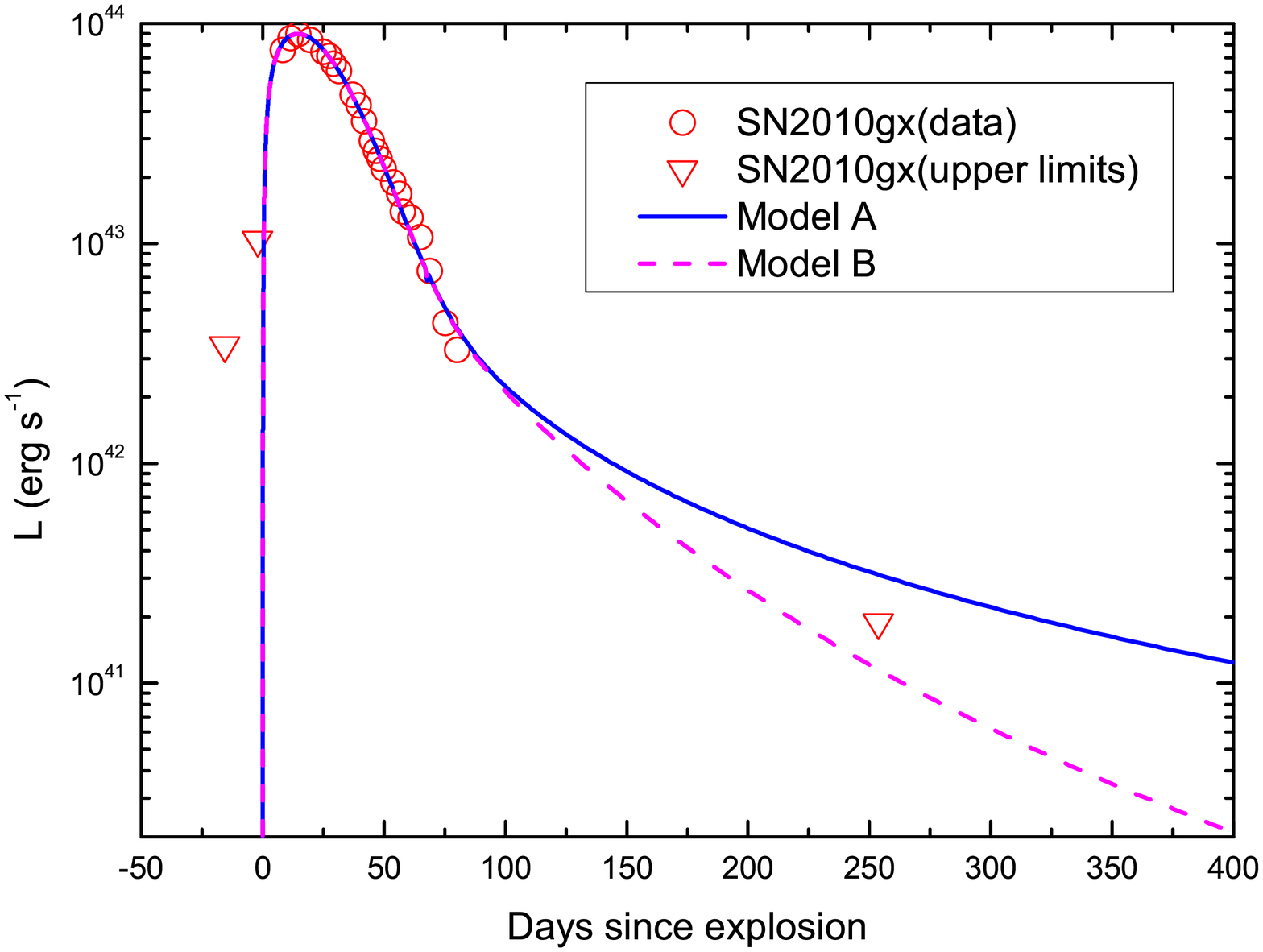}
   \caption{Light curves in the original magnetar-powered model and the revised magnetar-powered model for SN~2010gx.
The solid line and dashed line
 represent the cases that $A$~=~infinity and $2.212 \times 10^{14}$~s$^2$ respectively.
 Parameters for all the models are listed in Table \ref{parameters}.}
   \label{fig1}
\end{center}
\end{figure}

\begin{figure}[htbp]
\begin{center}
   \includegraphics[width=1.0\textwidth,angle=0]{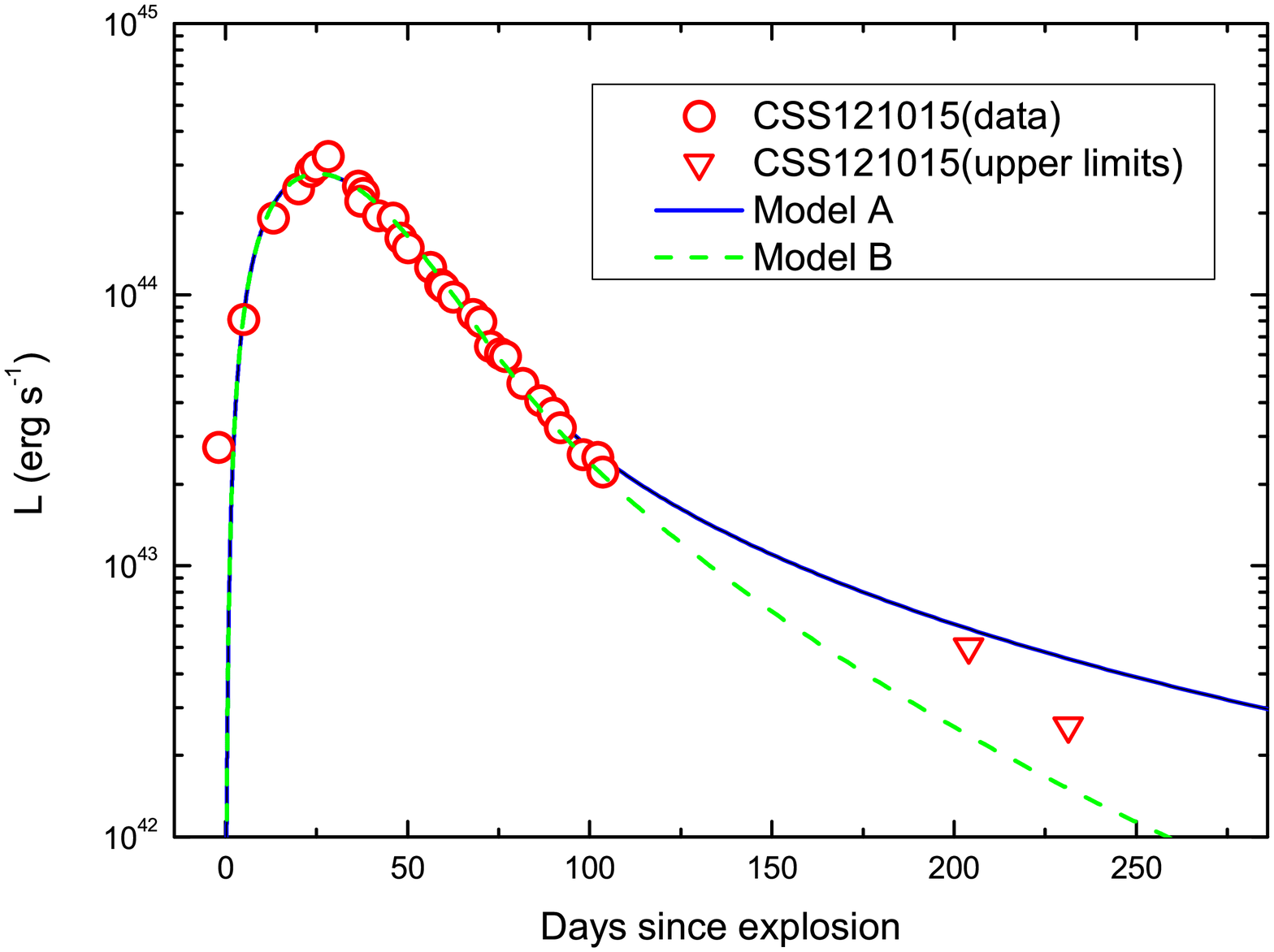}
   \caption{Light curves in the original magnetar-powered model and the revised magnetar-powered model for CSS121015.
The solid line and dashed line
 represent the cases that $A$~=~infinity and $1.612 \times 10^{14}$~s$^2$ respectively.
 Parameters for all the models are listed in Table \ref{parameters}.}
   \label{fig2}
\end{center}
\end{figure}

\begin{figure}[htbp]
\begin{center}
   \includegraphics[width=1.0\textwidth,angle=0]{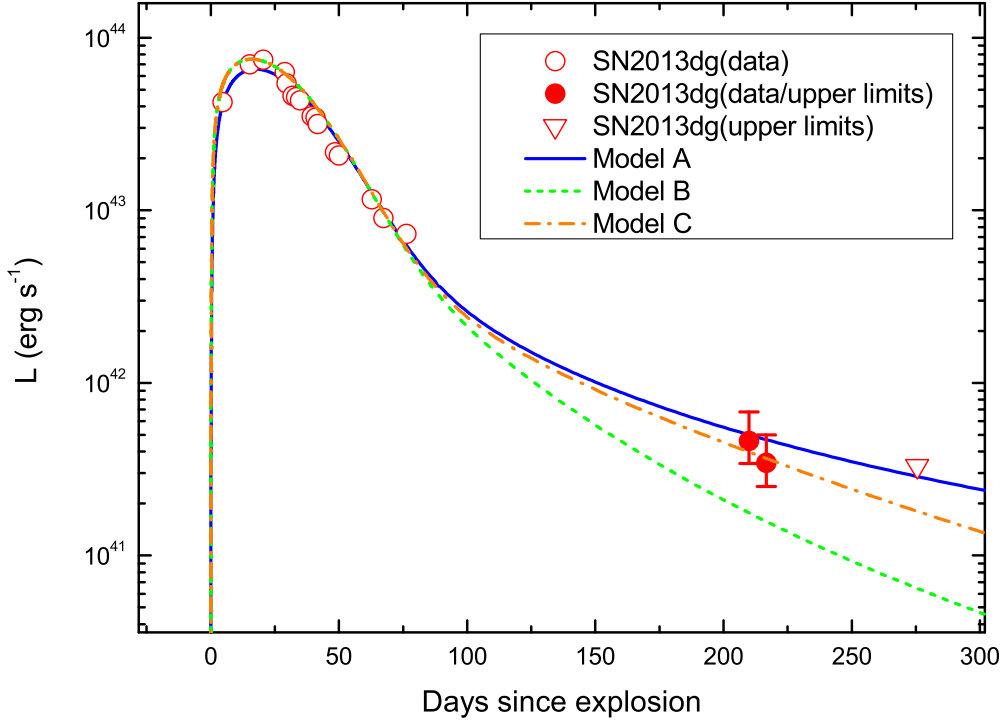}
   \caption{Light curves in the original magnetar-powered model and the revised magnetar-powered model for SN~2013dg.
The solid line, dotted line
and dash-dotted line
 represent the cases that $A$~=~infinity, $1.583 \times 10^{14}$~s$^2$ and $6.5 \times 10^{14}$~s$^2$, respectively.
 Data and parameters for Model A are obtained from \citet{Nich2014}.
 \textit{Note}: The contributions from the host galaxy have not been subtracted
 for last two data (at $\sim 200$~days). If the flux from SN~2013dg is comparable to the flux of the host galaxy,
 the late-time excess may still exist in the original magnetar-powered model.}
   \label{fig3}
\end{center}
\end{figure}

\begin{figure}[htbp]
\begin{center}
   \includegraphics[width=1.0\textwidth,angle=0]{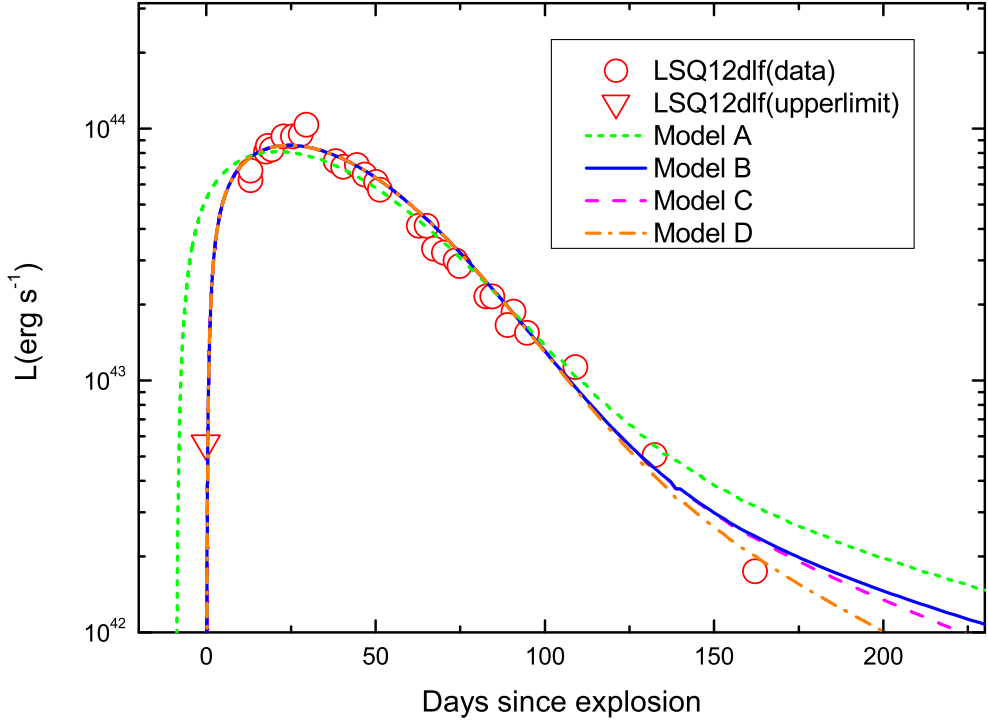}
   \caption{Light curves in the original magnetar-powered model and the revised magnetar-powered model for LSQ12dlf.
 The short dashed line represent the light curve reproduced by parameters adopted by \citet{Nich2014}.
In order to avoid the early time excess, we adopt other three sets of parameters, see Models B, C and D for LSQ12dlf in Table \ref{parameters}.
 The solid line, dashed line and dash-dotted line represent the cases that $A$~=~infinity, $7.62 \times 10^{14}$~s$^2$ and $3.50 \times 10^{14}$~s$^2$, respectively.
 Parameters for all the models are listed in Table \ref{parameters}.}
   \label{fig4}
\end{center}
\end{figure}

\begin{figure}[htbp]
\begin{center}
   \includegraphics[width=1.0\textwidth,angle=0]{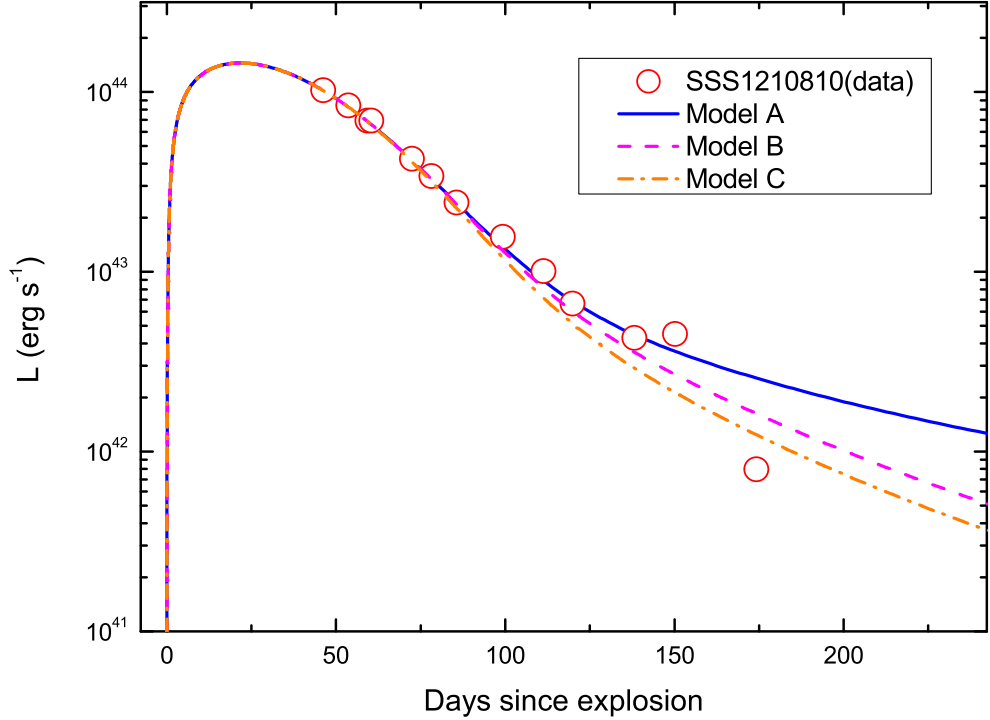}
   \caption{Light curves in the original magnetar-powered model and the revised magnetar-powered model for SSS1210810.
The solid line, dashed line
and dash-dotted line
 represent the cases that $A$~=~infinity, $2.28 \times 10^{14}$~s$^2$ and $1.50 \times 10^{14}$~s$^2$ respectively.
 Parameters for all the models are listed in Table \ref{parameters}.}
   \label{fig5}
\end{center}
\end{figure}

The fact that the late-time excess
can be eliminated or decreased by hard emission leakage indicates that it is non-trivial to introduce the leakage factor in the magnetar-powered model.
 The values of $A$ adopted
 in our sample are $\simeq 10^{14}-10^{15}$~s$^2$. When $t^{2}~\gtrsim A$, i.e., $t~\gtrsim 10^{7}-3 \times10^{7}$~s $\simeq 120-350$~days,
 the light curves reproduced by the magnetar-powered models with and without leakage can be easily distinguished from each other,
 see Figures \ref{fig1} - \ref{fig5}. For the same reason, the leakage effect does not significantly influence the early-time ($\lesssim 100-300$~days) light curves of these SLSNe.

When we take the limit $A~\rightarrow~\infty$, the leakage in the ejecta can be neglected, i.e., all gamma ray and X-ray photons should be trapped and thermalized to UV-optical-NIR photons. The limit $A~\rightarrow~\infty$ might be a good approximation for some other SLSNe-Ic, e.g., PTF10hgi, SN~2011kf, and SN~2012il, but we have no idea whether PTF11rks needs to be explained by the revised magnetar-powered model, due to the lacking of the later time ($\gtrsim 100$~days post explosion) data and strict upper limits, see Figure 12 in \citet{Inse2013}. The possibility that the late-time leakage can be compensated by $^{56}$Co decay tails cannot be
 excluded and it might be worthwhile to verify this possibility in future modelings. Another possibility accounts for full trapping is that the gamma ray and X-ray photons are so soft that $\kappa_{\gamma}$ ($\kappa_{\gamma/X}$) is very large and the values of $A$ far exceed $10^{15}$~s$^2$.

Whether the leakage of hard emissions can plays a key role in later phases
is of great interest and of great importance in future studies.
A larger sample of SLSNe with high
quality observed data, especially recorded after the SLSNe transitioned to
the nebular phases ($\gtrsim 60-100$~days after the explosions), is required to determine the precise values of
fading rate at nebular phases and therefore precise values of the parameter $A$ in Equation (\ref{equ:L-mag}).

Finally, as addressed by \citet{Des2012,Des2013}, theoretical light curves of SLSNe alone are inadequate to unveil the nature of SLSNe.
To completely clarify the energy generating mechanisms, both theoretical light curves and spectra for every individual event are needed to be examined carefully.
Further detailed modelings are also needed to determine precisely the parameters of putative nascent magnetars.

\acknowledgments
We thank an anonymous referee for helpful comments and suggestions that have allowed us to improve this manuscript.
This work is supported by the National Basic Research Program (``973" Program)
of China (Grants 2014CB845800 and 2013CB834900) and the National Natural Science Foundation of China (grants Nos. 11033002 and 11322328). XFW was also partially supported by the One-Hundred-Talent Program, the Youth Innovation Promotion Association, and the Strategic Priority Research Program
``The Emergence of Cosmological Structures" (Grant No. XDB09000000) of
the Chinese Academy of Sciences, and the Natural Science Foundation of Jiangsu Province.

\clearpage
\begin{table*}
\begin{center}
\caption{Fit parameters}\label{parameters}
\begin{tabular}{ccccccccccc}
\hline
\hline
 & $M_{\rm ej}$  & $B$           & $P$  & $v$/$c$ $^{\dag}$ & $\kappa$          & $\eta$ $^{\ddag}$ & $A$     & Reference \\
 \hline
 & ($M_{\odot}$) & ($10^{14}$~G) & (ms) &                   & (cm$^2$ g$^{-1}$) &                   & (s$^2$) &  \\
\hline
\hline

{\bf SN~2010gx}\\
\hline
Model A & 7.1 & 7.4 & 2.0 & 0.0762 & 0.15 & 0 &        $\infty $       & \citet{Inse2013}\\
Model B & 7.1 & 7.4 & 2.0 & 0.0762 & 0.15 & 0 & $2.212 \times 10^{14}$ &  This work      \\
\hline
\hline

{\bf CSS121015}\\
\hline
Model A & 5.5 & 2.1 & 2.0 & 0.06 & 0.2 & 0 &        $\infty $         & \citet{Nich2014}\\
Model B & 5.5 & 2.1 & 2.0 & 0.06 & 0.2 & 0 & $1.612 \times 10^{14}$   &  This work      \\
\hline
\hline

{\bf SN~2013dg}\\
\hline
Model A & 5.4 & 7.1 & 2.5 & 0.06 & 0.2  & 0    &        $\infty $         & \citet{Nich2014}\\
Model B & 5.4 & 7.4 & 1.9 & 0.06 & 0.2  & 0    & $1.583 \times 10^{14}$   & This work       \\
Model C & 5.4 & 7.4 & 1.9 & 0.06 & 0.2  & $>$0 & $6.5   \times 10^{14}$   & This work       \\
\hline
\hline

{\bf LSQ12dlf}\\
\hline
Model A & 10.0 & 3.7 & 1.9 & 0.03 & 0.2   & 0 & $\infty $               & \citet{Nich2014}\\
Model B & 10.0 & 4.5 & 1.4 & 0.03 & 0.13  & 0 & $\infty $              & This work       \\
Model C & 10.0 & 4.5 & 1.4 & 0.03 & 0.13  & 0 & $7.62 \times 10^{14}$  & This work       \\
Model D & 10.0 & 4.5 & 1.4 & 0.03 & 0.13  & $>$0 & $3.50 \times 10^{14}$  & This work       \\
\hline
\hline

{\bf SSS120810}\\
\hline
Model A & 12.5 & 3.9 & 1.2 & 0.076 & 0.2  & 0 & $\infty $              & \citet{Nich2014}\\
Model B & 12.5 & 3.9 & 1.2 & 0.076 & 0.2  & 0 & $2.28 \times 10^{14}$  & This work       \\
Model C & 12.5 & 3.9 & 1.2 & 0.076 & 0.2  & $>$0 & $1.50 \times 10^{14}$  & This work       \\
\hline
\end{tabular}
\end{center}
$^{\dag}$ The values of the expansion velocities of the ejecta have not been listed by \citet{Inse2013} and \citet{Nich2014}. \\
$^{\ddag}$ The density of the ejecta can be represented by $\rho(r, t)\propto [r(t)]^{-\eta}$. \\
\end{table*}

\clearpage

\clearpage

\clearpage

\clearpage

\clearpage


\begin{thebibliography}{}

\bibitem[Arnett(1982)]{Arn1982}{Arnett}, W.~D. 1982, \apj, 253, 785
\bibitem[Baltay(2013)]{Bal2013}{Baltay}, C., {Rabinowitz}, D., {Hadjiyska}, E., {et~al.} 2013, \pasp, 125, 683
\bibitem[Barkat et~al.(1967)]{Bar1967}{Barkat}, Z., {Rakavy}, G., \& {Sack}, N. 1967, \prl, 18, 379
\bibitem[Barkov \& Komissarov(2011)]{Bar2011}{Barkov}, M.~V., \& {Komissarov}, S.~S. 2011, \mnras, 415, 944
\bibitem[Ben-Ami et~al.(2014)]{Ben-A2014}{Ben-Ami}, S., {Gal-Yam}, A., {Mazzali}, P.~A., {et~al.} 2014, \apj, 785, 37
\bibitem[Benetti et~al.(2014)]{Ben2014}{Benetti}, S., {Nicholl}, M., {Cappellaro}, E., {et~al.} 2014, \mnras, 441, 289
\bibitem[Bucciantini et~al.(2008)]{Buc2008}{Bucciantini}, N., {Quataert}, E., {Arons}, J., {et~al.} 2008, \mnras, 383, L25.
\bibitem[Chatzopoulos et~al.(2009)]{Cha2009}{Chatzopoulos}, E., {Wheeler}, J.~C., \& {Vinko}, J. 2009, \apj, 704, 1251
\bibitem[Chatzopoulos et~al.(2012)]{Cha2012}{Chatzopoulos}, E., {Wheeler}, J.~C., \& {Vinko}, J. 2012, \apj, 746, 121
\bibitem[Chatzopoulos \& Wheeler(2012a)]{Cha2012a}{Chatzopoulos}, E., \& {Wheeler}, J.~C. 2012a, \apj, 748, 42
\bibitem[Chatzopoulos \& Wheeler(2012b)]{Cha2012b}{Chatzopoulos}, E., \& {Wheeler}, J.~C. 2012b, \apj, 760, 154
\bibitem[Chatzopoulos et~al.(2013a)]{Cha2013a}{Chatzopoulos}, E., {Wheeler}, J.~C., {Vinko}, J., {et~al.} 2013a, \apj, 773, 76
\bibitem[Chatzopoulos et~al.(2013b)]{Cha2013b}{Chatzopoulos}, E., {Wheeler}, J.~C., \& {Couch}, S.~M. 2013, \apj, 776, 129
\bibitem[Chen et~al.(2014a)]{Che2014a}{Chen}, K.-J., {Woosley}, S., {Heger}, A., {et~al.} 2014a, \apj, 792, 28
\bibitem[Chen et~al.(2014b)]{Che2014b}{Chen}, K.-J., {Heger}, A., {Woosley}, S., {et~al.} 2014b, \apj, 792, 44
\bibitem[Chen et~al.(2014c)]{Che2014c}{Chen}, K.-J., {Heger}, A., {Woosley}, S., {et~al.} 2014c, arXiv:1407.7550
\bibitem[Chen et~al.(2014)]{Che2014}{Chen}, T.-W., {Smartt}, S.~J., {Jerkstrand}, A., {et~al.} 2014, arXiv:1409.7728
\bibitem[Chevalier(1982)]{Che1982}{Chevalier}, R.~A. 1982, \apj, 258, 790
\bibitem[Chevalier \& Fransson(1994)]{Che1994}{Chevalier}, R.~A., \& {Fransson}, C. 1994, \apj, 420, 268
\bibitem[Chevalier \& Irwin(2011)]{Che2011}{Chevalier}, R.~A., \& {Irwin}, C.~M. 2011, \apjl, 729, L6
\bibitem[Chugai \& Danziger(1994)]{Chu1994}{Chugai}, N.~N., \& {Danziger}, I.~J. 1994, \mnras, 268, 173
\bibitem[Chugai(2009)]{Chu2009}{Chugai}, N.~N. 2009, \mnras, 400, 866
\bibitem[Churazov et~al.(2014)]{Chu2014}{Churazov}, E., {Sunyaev}, R., {Isern} J., {et~al.} 2014, \nat, 512, 406
\bibitem[Clocchiatti \& Wheeler(1997)]{Clo1997}{Clocchiatti}, A., \& {Wheeler}, J.~C. 1997, \apj, 491, 375
\bibitem[Colgate \& McKee(1969)]{Col1969}{Colgate}, S.~A., \& {McKee}, C. 1969, \apj, 157, 623
\bibitem[Colgate et~al.(1980)]{Col1980}{Colgate}, S.~A., {Petschek}, A.~G., \& {Kriese}, J.~T. 1980, \apjl, 237, L81
\bibitem[Dai(2004)]{Dai2004}{Dai}, Z.~G. 2004, \apj, 606, 1000
\bibitem[Dai \& Lu(1998a)]{Dai1998a}{Dai}, Z.~G., \& {Lu}, T. 1998a, \aap, 333, L87
\bibitem[Dai \& Lu(1998b)]{Dai1998b}{Dai}, Z.~G., \& {Lu}, T. 1998b, \prl, 81, 4301
\bibitem[Dai \& Liu(2012)]{Dai2012}{Dai}, Z.~G., \& {Liu}, R.~Y. 2012, \apj, 759, 58
\bibitem[Dall'Osso et~al.(2011)]{Dal2011}{Dall'Osso}, S., {Stratta}, G., {Guetta}, D., {et~al.} 2011, \aap, 526, 121
\bibitem[Dessart et~al.(2012)]{Des2012}{Dessart}, L., {Hillier}, D.~J., {Waldman}, R., {et~al.} 2012, \mnras, 426, L76
\bibitem[Dessart et~al.(2013)]{Des2013}{Dessart}, L., {Waldman}, R., {Livne}, E., {et~al.} 2013, \mnras, 428, 3227
\bibitem[Diehl et~al.(2014)]{Dieh2014}{Diehl}, R., {Siegert}, T., {Hillebrandt}, W., {et~al.} 2014, arXiv:1409.5477
\bibitem[Diehl et~al.(2014)]{Dra2009}{Drake}, A.~J., {Djorgovski}, S.~G., {Mahabal}, A., {et~al.} \apj, 696, 870
\bibitem[Gal-Yam et~al.(2009)]{Gal2009}{Gal-Yam}, A., {Mazzali}, P., {Ofek}, E. O., {et~al.} 2009, \nat, 462, 624
\bibitem[Gal-Yam(2012)]{Gal2012}{Gal-Yam}, A. 2012, Science, 337, 927
\bibitem[Heger \& Woosley(2003)]{Heg2002}{Heger}, A., \& {Woosley}, S.~E. 2002, \apj, 567, 532
\bibitem[Heger et~al.(2003)]{Heg2003}{Heger}, A., {Fryer}, C.~L., {Woosley}, S.~E., {et~al.} 2003, \apj, 591, 288
\bibitem[Inserra et~al.(2013)]{Inse2013}{Inserra}, C., {Smartt}, S.~J., {Jerkstrand}, A., {et~al.} 2013, \apj, 770, 128
\bibitem[Kaiser(2010)]{Kai2010}{Kaiser}, N., {Burgett}, W., {Chambers}, K., {et~al.} 2010, in Ground-based and Airborne Telescopes III, Vol. 7733, Society of
    Photo-Optical Instrumentation Engineers (SPIE) Conference Series, 77330
\bibitem[Kasen \& Bildsten(2010)]{Kas2010}{Kasen}, D., \& {Bildsten}, L. 2010, \apj, 717, 245
\bibitem[Kleiser \& Kasen(2014)]{Kle2014}{Kleiser}, Io K.~W., \& {Kasen}, D. 2014, \mnras, 438, 318
\bibitem[Komissarov \& Barkov(2008)]{Kom2008}{Komissarov}, S.~S., \& {Barkov}, M.~V. 2008, \mnras, 382, 1029
\bibitem[Kotera et~al.(2013)]{Kot2013}{Kotera}, K., {Phinney}, E.~S., \& {Olinto}, A.~V. 2013, \mnras, 432, 3228
\bibitem[Kouveliotou et~al.(1994)]{Kou1994}{Kouveliotou}, C., {Fishman}, G.~J., {Meegan}, C.~A., {et~al.} 1994, \nat, 368, 125
\bibitem[Kouveliotou et~al.(1998)]{Kou1998}{Kouveliotou}, C., {Dieters}, S., {Strohmayer}, T., {et~al.} 1998, \nat, 393, 235
\bibitem[Kozyreva et~al.(2014)]{Koz2014}{Kozyreva}, A., {Yoon}, S.-C., \& {Langer}, N. 2014, \aap, 566, 146
\bibitem[Kudritzki et~al.(1987)]{Kud1987}{Kudritzki}, R.~P., {Pauldrach}, A.~W.~A., \& {Puls}, J. 1987, \aap, 173, 293
\bibitem[Law(2009)]{Law2009}{Law}, N.~M., {Kulkarni}, S.~R., {Dekany}, R.~G., {et~al.} 2009, \pasp, 121, 1395
\bibitem[Leitherer et~al.(1992)]{Lei1992}{Leitherer}, C., {Robert}, C., \& {Drissen}, L. 1992, \apj, 401, 596
\bibitem[Lyne et~al.(1998)]{Lyn1998}{Lyne}, A.~G., {Manchester}, R.~N., {Lorimer}, D.~R., {et~al.} 1998, \mnras, 295, 743
\bibitem[Maeda et~al.(2007)]{Mae2007}{Maeda}, K., {Tanaka}, M., {Nomoto}, K., {et~al.} 2007, \apj, 666, 1069
\bibitem[Matz et~al.(1988)]{Mat1988}{Matz}, S.~M., {Share}, G.~H., {Leising}, M.~D., {et~al.} 1988, \nat, 331, 416
\bibitem[Metzger et~al.(2007)]{Met2007}{Metzger}, B.~D., {Thompson}, T.~A., \& {Quataert}, E. 2007, \apj, 659, 561
\bibitem[Metzger et~al.(2011)]{Met2011}{Metzger}, B.~D., {Giannios}, D., {Thompson}, T.~A., {et~al.} 2011, \mnras, 413, 2031
\bibitem[Metzger et~al.(2014)]{Met2014}{Metzger}, B.~D., {Vurm}, I., {Hasco\"et}, R., {et~al.} 2014, \mnras, 437, 703
\bibitem[Moriya et~al.(2010)]{Mor2010}{Moriya}, T., {Tominaga}, N., {Tanaka}, M., {et~al.} 2010, \apjl, 717, L83
\bibitem[Nicholl et~al.(2013)]{Nich2013}{Nicholl}, M., {Smartt}, S.~J., {Jerkstrand}, A., {et~al.} 2013, \nat, 502, 346
\bibitem[Nicholl et~al.(2014)]{Nich2014}{Nicholl}, M., {Jerkstrand}, A., {Inserra}, C., {et~al.} 2014, \mnras, 444, 2096
\bibitem[Pastorello et~al.(2008)]{Pas2008}{Pastorello}, A., {Mattila}, S., {Zampieri}, L., {et~al.} 2008, \mnras, 389, 113
\bibitem[Quimby(2012)]{Qui2012}{Quimby}, R.~M. 2012, IAUS, 279, 22
\bibitem[Rakavy \& Shaviv(1967)]{Rak1967}{Rakavy}, G., \& {Shaviv}, G. 1967, \apj, 148, 803
\bibitem[Rau(2009)]{Rau2009}{Rau}, A., {Kulkarni}, S.~R., {Law}, N.~M., {et~al.} 2009, \pasp, 121, 1334
\bibitem[Sollerman et~al.(2000)]{Sol2000}{Sollerman}, J., {Kozma}, C., {Fransson}, C., {et~al.} 2000, \apjl, 537, L127
\bibitem[Sollerman et~al.(2002)]{Sol2002}{Sollerman}, J., {Holland}, S.~T., {Challis}, P., {et~al.} \aap, 386, 944
\bibitem[Sunyaev et~al.(1987)]{Sun1987}{Sunyaev}, R., {Kaniovsky}, A., {Efremov}, V., {et~al.} 1987, \nat, 330, 227
\bibitem[Tonry(2012)]{Ton2012}{Tonry}, J.~L., {Stubbs}, C.~W., {Lykke}, K.~R., {et~al.} 2012, \apj, 745, 42
\bibitem[van Paradijs(1995)]{Par1995}{van Paradijs}, J., {Taam}, R.~E., \& {van den Heuvel}, E.~P.~J. 1995, \aap, 299, L41
\bibitem[Valenti et~al.(2008)]{Val2008}{Valenti}, S., {Benetti}, S., {Cappellaro}, E., {et~al.} 2008, \mnras, 383, 1485
\bibitem[Vink et~al.(2001)]{Vin2001}{Vink}, J., {de Koter}, A., \& {Lamers}, H.~J.~G.~L.~M. 2001, \aap, 369, 574
\bibitem[Vink \& Kuiper(2006)]{Vin2006}{Vink}, J., \& {Kuiper}, L. 2006, \mnras, 370, L14
\bibitem[Whalen et~al.(2013)]{Wha2013}{Whalen}, D., {Even}, W., {Frey}, L., {et~al.} 2013, \apj, 777, 110
\bibitem[Woosley et~al.(1989)]{Woos1989}{Woosley}, S., {Hartmann}, D., \& {Pinto}, P.~A. 1989, \apj, 346, 395
\bibitem[Woosley et~al.(2007)]{Woos2007}{Woosley}, S.~E., {Blinnikov}, S., \& {Heger}, A. 2007, \nat, 450, 390
\bibitem[Woosley(2010)]{Woos2010}{Woosley}, S.~E. 2010, \apjl, 719, L204
\bibitem[Yu \& Dai(2007)]{Yu2007}{Yu}, Y.~W., \& {Dai}, Z.~G. 2007, \aap, 470, 119
\bibitem[Zhang(2007)]{Zhang2007}{Zhang}, B. 2007, Chin. J. Astron. Astrophys., 7, 1
\bibitem[Zhang \& M{\'e}sz{\'a}ros(2001)]{Zhang2001}{Zhang}, B., \& {M{\'e}sz{\'a}ros}, P. 2001, \apjl, 552, L35

\end{thebibliography}
\end{document}